\documentclass[conference]{IEEEtran}
\IEEEoverridecommandlockouts
\usepackage{cite}
\usepackage{amsmath,amssymb,amsfonts}
\usepackage{algorithmic}
\usepackage{graphicx}
\usepackage{textcomp}
\usepackage{xcolor}
\usepackage{url}
\def\BibTeX{{\rm B\kern-.05em{\sc i\kern-.025em b}\kern-.08em
    T\kern-.1667em\lower.7ex\hbox{E}\kern-.125emX}}
\begin{document}

\newcommand{\sebastian}[1]{\textbf{\textcolor{green}{[S: #1]}}}
\newcommand{\karolina}[1]{\textcolor{purple}{K: #1}}
\newcommand{\gloria}[1]{\textcolor{orange}{G: #1}}
\newcommand{\vesa}[1]{\textcolor{red}{V: #1}}
\title{SIMILARITY METRICS FOR LATE REVERBERATION\\
\thanks{The work of the first author was funded by the Aalto University School of Electrical Engineering.}
}
\author{\IEEEauthorblockN{Gloria Dal Santo$^1$, Karolina Prawda$^1$, Sebastian J. Schlecht$^2$, Vesa Välimäki$^1$}
\IEEEauthorblockA{$^1$ \textit{Acoustics Lab, Department of Information and Communications Engineering},\\ Aalto University, Finland\\
$^2$ \textit{Multimedia Communications and Signal Processing},\\ 
Friedrich-Alexander-Universität Erlangen-Nürnberg (FAU), Erlangen, Germany}}

\maketitle

\begin{abstract}
% Motivation - Problem statement - Approach - Results - Conclusion 
Automatic tuning of reverberation algorithms relies on the optimization of a cost function. While general audio similarity metrics are useful, they are not optimized for the specific statistical properties of reverberation in rooms. This paper presents two novel metrics for assessing the similarity of late reverberation in room impulse responses. These metrics are differentiable and can be utilized within a machine-learning framework. 
We compare the performance of these metrics to two popular audio metrics using a large dataset of room impulse responses encompassing various room configurations and microphone positions. The results indicate that the proposed functions based on averaged power and frequency-band energy decay outperform the baselines with the former exhibiting the most suitable profile towards the minimum. The proposed work holds promise as an improvement to the design and evaluation of reverberation similarity metrics.
\end{abstract}

% ================= INTRO ================= %
\section{Introduction}\label{sec:intro}
A room impulse response (RIR) describes sound propagation in an enclosed space, playing an important role in acoustic analysis. Quantities derived from the RIR, such as the energy decay curve (EDC) \cite{schroeder1965new} and reverberation time (RT), provide concise descriptions of a room's sound field. 
A typical RIR can be conceptualized as comprising three distinct stages: direct sound, early reflections, and late reverberation, with the latter being the focus of this study.

Late reverberation occurs when an increased number of superposed reflections makes individual reflections indistinguishable from the auditory system. During this stage, the sound field becomes diffuse and is best described by its statistical properties \cite{schroeder1961natural}. A diffuse sound field is then considered homogeneous and isotropic \cite{kuttruff2016room, waterhouse1955interference}. The evolving power spectrum, particularly its decay rate, indicates the size of the space and the absorption properties of its materials \cite{beranek2006analysis}. Due to these characteristics, and disregarding the filtering effect by the absorption properties of the room and medium, late reverberation is frequently associated with exponentially decaying white noise \cite{moorer1979reverberation}.

Artificial reverberation encompasses various techniques and algorithms designed to replicate the acoustic characteristics of specific environments \cite{valimaki2012fifty}. However, tuning the parameters of an artificial reverberation algorithm to match a target RIR perceptually is non-trivial. In literature, several approaches to automatic parameter tuning have been proposed, including genetic algorithms \cite{coggin2016automatic, chemistruck2012generating, shen2020data, ibnyahya2022method}, and stochastic gradient descent \cite{dal2023differentiable, santo2024feedback}. 

Recently proposed neural networks for RIR estimation typically comprise an encoder for feature extraction and a generator for synthesis of RIRs from these features \cite{steinmetz2021filtered, lee2022differentiable, ratnarajah2023towards, lee2023yet}. The performance of both approaches heavily relies on the choice of the cost function, for which two main trends can be identified: one utilizes metrics based on acoustic quantities, such as EDC \cite{coggin2016automatic, ratnarajah2023towards}, echo density \cite{mezza2024data}, and RT \cite{shen2020data}; the other relies on element-wise distances of spectrograms, including the short-time Fourier transform (STFT) \cite{steinmetz2021filtered, lee2022differentiable, lee2023yet} and mel-frequency cepstral coefficients \cite{coggin2016automatic, ibnyahya2022method}. 

Following the first trend of acoustic metrics, Helmholz et al.~proposed a prediction model\cite{helmholz2022towards}, in which a combination of standard acoustic parameters \cite{iso2009measurement, zahorik2009perceptually} was designed based on a subjective listening test to predict the perceived RIR similarity. However, in machine learning (ML) applications for automatic tuning of reverberation algorithms, computing some of these quantities is impractical. Moreover, the estimation techniques used to derive the acoustic parameters are subject to uncertainties, which are then propagated to the estimated values. On the other hand, time-frequency representations, while effective in many synthesis tasks, do not fully exploit the statistical character of RIRs. 

Quantifying the performance of a similarity metric itself poses a highly non-trivial challenge. To establish subjective correlations, listening tests with a large number of subjects and test configurations are essential. Many RIR datasets lack metadata, including absorption coefficients, room geometry, and the locations of transducers within the space. This lack of information impedes the comprehensive coverage of reverberation conditions and degrees of dissimilarities, resulting in gaps during evaluation.

In this paper, we present two novel similarity metrics for RIRs and compare them with two metrics commonly used in audio synthesis tasks, namely the multi-scale spectral loss (MSS) \cite{yamamoto2020parallel} and the error-to-signal ratio (ESR) \cite{wright2019real}. Our study focuses on the similarity between the late reverberation of RIRs recorded in a room with variable acoustics. By adopting this approach, we aim to establish a correlation between the metrics and the specific characteristics of the measurement setup, rather than solely relying on acoustic parameters.

The paper is organized as follows. Sec.~\ref{sec:method} presents the similarity metrics proposed in this study. The evaluation setup and results are described in Sec.~\ref{sec:eval}, followed by a discussion on the outcomes in Sec.~\ref{sec:discussion}. Sec.~\ref{sec:conclusion} offers concluding remarks.
% ================= METHOD ================= %
\section{Proposed late reverberation similarity metrics}\label{sec:method}
Unlike early reflections, late reverberation's statistical properties in a room are more predictable and consistent across various locations \cite{schroeder1961natural}. Also, for diffusing rooms, location-dependent features are often hard to perceive for the late reverberation \cite{mckenzie2023role}. We propose a new similarity metric leveraging these features, using local signal power averages. In addition, we introduce a frequency-dependent EDC distance. 

In the following, we assume that the direct sound and early reflections have been removed from RIRs. Therefore, the late reverberation is considered to start at time $t=0$. % Both metrics are differentiable, making them suitable for integration into machine learning frameworks.
\subsection{Averaged power convergence}
When comparing the late reverberation of two RIRs, the similarity to exponentially decaying white Gaussian noise can lead to noisy values in sample-to-sample distances. Averaging across multiple time-frequency bins smooths out short-term fluctuations, leading to more reliable distance predictors. 

Building upon this premise, we propose a novel similarity metric between a target, $h(t)$, and an analyzed RIR, $\hat{h}(t)$. This metric, called averaged power convergence (PC), is based on local time-frequency signal power averages and computed as
%\begin{align}
 %   \mathcal{L}_{\textrm{PC}} = \frac{\lVert\lvert H(t, f)\rvert^2*W  - \lvert \hat{H}(t, f) \rvert^2*W\rVert_{\textrm{F}}}{\lVert\lvert H(t, f) \rvert^2*W\rVert_{\textrm{F}} \, \lVert\lvert \hat{H}(t, f) \rvert^2*W
  %  \rVert_{\textrm{F}}}\,,
%\end{align}
\begin{align}
    \mathcal{L}_{\textrm{PC}} &= \left\lVert \frac{| H(t, f)|^2 * W  - |\hat{H}(t, f) |^2 * W}{(| H(t, f) |^2 * W) (| \hat{H}(t, f) |^2 * W)} \right\rVert_{\textrm{F}}\,,
\end{align}
where $\lvert H(t, f) \rvert^2$ is the squared magnitude STFT of $h(t)$, $W$ is a time-frequency Hann window, $\lVert\cdot\rVert_\textrm{F}$ is the Frobenius
norm, and $*$ denotes the 2D convolution operation, in the deep-learning sense, with a non-unitary stride. By using the convolution operation, the loss emphasizes differences in the local time-frequency-averaged power of the magnitude STFT, assumed to converge to zero for two RIRs measured in the same reverberation conditions. In this work, to compute the STFT, we used a window length of 1024 samples, with $25\%$ hop size. The Hann window $W$ is a $64 \times 64 $ matrix applied with a symmetric stride of 4.   % \gloria{TODO: add something about the influence of the window size in the accuracy of the variance estimator  }
\subsection{Energy decay convergence}
The EDC is a descriptor of the level of energy over time, which is used to calculate the RT in RIRs \cite{schroeder1965new}. Given $h(t)$ of length $L$, the EDC is computed through Schroeder backward integration: 
\begin{align}
    \varepsilon(t; f_\textrm{c}) = \sum_{\tau=t}^Lh_{f_\textrm{c}}^2(\tau)\,,
\end{align}
where $h_{f_\textrm{c}}$ is the input RIR at a frequency band with center frequency $f_\textrm{c}$. It is usually reported on a decibel scale, here denoted by $\varepsilon_{\textrm{dB}}$. 
In line with the loss functions presented in \cite{mezza2024data, ratnarajah2022mesh2ir}, we propose a similarity metric on $\varepsilon_{\textrm{dB}}$ computed as
\begin{align}\label{eq:edc_loss}
    \mathcal{L}_{\text{EDC}} = \frac{1}{|\mathcal{C}|}\sum_{f_{\textrm{c}} \in \mathcal{C}}\frac{\sum_{t=0}^L ( \varepsilon_{\textrm{dB}}(t; f_\textrm{c}) - \hat{\varepsilon}_{\textrm{dB}}(t; f_\textrm{c}))^2}{\sum_{t=0}^L \varepsilon_{\textrm{dB}}^2(t; f_\textrm{c})}\,,
\end{align}
where $|\cdot|$ indicates the cardinality of a set. As opposed to \cite{mezza2024data, ratnarajah2022mesh2ir}, we average the EDC computed on a set $\mathcal{C}$ of 29 one-third octave bands ranging from 20 Hz to 12.5 kHz. 

When using backward integration, background noise affects the entire EDC, leading to a vertical displacement at the beginning of the EDC \cite{karjalainen2002}. To avoid emphasizing differences in noise level, all EDCs are normalized to 0 dB prior to computing \eqref{eq:edc_loss}.

% ================= EVALUATION ================= %
\section{Objective Evaluation}\label{sec:eval}
\begin{figure*}[ht!]
\centerline{\includegraphics[trim={4.5cm 2cm 3cm 4cm},width=\textwidth]{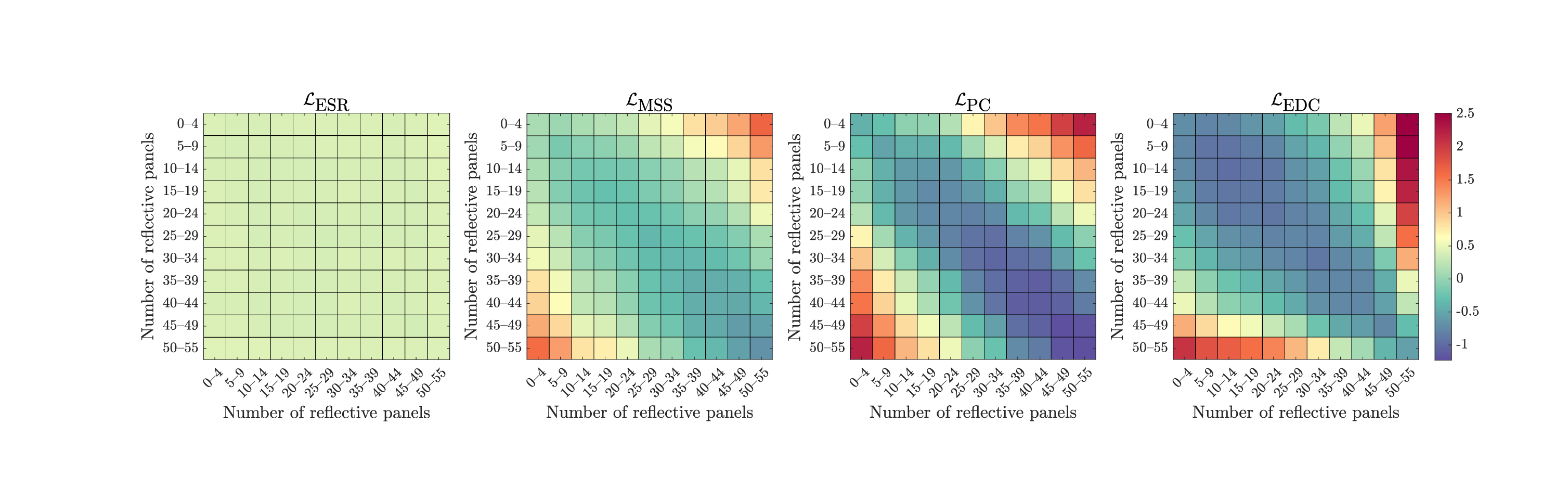}}
% \vspace{-1mm}
\caption{\textit{Median values of the standardized similarity metric distribution for each pair of reflective panel conditions. From left to right: ESR, MSS loss, averaged PC, EDC convergence. Labels of the Y-axis refer to the reference RIR $h(t)$, while X-axis refer to the analyzed RIR $\hat{h}(t)$.}}
\label{fig:med_loss_nclosed}
\end{figure*}
We assess our proposed metrics against two losses commonly used in ML audio synthesis tasks. This section first discusses the evaluation dataset and the baseline functions. Following that, we describe the evaluation setup and provide an overview of the results. 
\subsection{Evaluation dataset}
In this study, we use a dataset of RIRs collected in the variable acoustics laboratory \emph{Arni} at Acoustics Lab of Aalto University, Espoo, Finland \cite{prawda2022calibrating}. 
%\emph{Arni} is a rectangular room with dimensions 8.9 m $\times$ 6.3 $\times$  3.6 m (length, width, and height, respectively). 
The walls and ceiling of \emph{Arni} are covered with 55 variable acoustics panels made from painted metal sheets and filled with absorptive material. The dataset contains RIRs from 5342 panel configurations and 5 microphone positions. 
The sound field in \emph{Arni} is assumed not to be fully diffuse, as unevenly distributed absorption \cite{stephenson2016rigorous} and the shoebox shape of \emph{Arni} \cite{badeau2019general} both indicate a lack of isotropy and homogeneity. Nonetheless, it assumes convergence of the statistical properties of the late reverberation for RIRs sharing similar room absorption configurations and microphone positions. One of the main advantages inherent to this dataset is its fine resolution, which results in smooth transitions between different reverberation conditions.

Since our focus lies solely on late reverberation, we remove the direct and early reflections from all analyzed RIRs. To detect the onset, we analyze the energy variation over time using the STFT to identify the frame with the most significant energy change. The onset time is then determined from the index of the STFT window after conversion to the time domain.

The mixing time $t_\textrm{mix}$ refers to the point in time beyond which the auditory system cannot differentiate between successive reflections \cite{blauert1997spatial}, delineating the transition between early reflections and late reverberation. We use a common value for $t_\textrm{mix}$, chosen as the maximum of the median mixing time among the configurations grouped by the number of reflective panels, which corresponds to the setup with 52 panels in reflective position.
More information about the data pre-processing is available online\,\footnotemark[1]. 

\footnotetext[1]{{\protect{\url{http://research.spa.aalto.fi/publications/papers/asilomar24-reverb-similarity}}}}

\subsection{Baselines}
The proposed losses are compared to the multi-scale spectral loss, MSS, a metric utilized within differentiable digital signal processing (DDSP) \cite{engel2020ddsp, hayes2024review}. MSS has gained attention in various audio synthesis applications, including areas related to reverb \cite{steinmetz2021filtered, lee2022differentiable, bona2022automatic, su2020acoustic}. MSS addresses the inherent trade-off between time-frequency resolution present in magnitude spectrograms by incorporating multiple STFTs with varying time-frequency resolutions into a unified loss function \cite{yamamoto2020parallel}. However, MSS suffers from instabilities when dealing with time shifts and nonstationarity behaviors in signals \cite{vahidi2023mesostructures}, which are typically assumed to be minimal when analyzing late reverberation. 

The MSS is composed of a spectral convergence term $\mathcal{L}_{\text{SC}}$ and a spectral log-magnitude term $\mathcal{L}_{\text{SM}}$, respectively:
\begin{align}    
\mathcal{L}_{\text{SC}}(h, \hat{h}) &= \frac{\lVert|{H}(t,f)|-|{\hat{H}(t,f)}|\rVert_F}{\lVert|{H}(t,f)|\rVert_F}
\end{align}
and
\begin{align}
\mathcal{L}_{\text{SM}}(h, \hat{h}) = \frac{1}{N} \lVert&\log(|{H}(t,f)|)-\log(|{\hat{H}(t,f)}|)\rVert_1\,,
\end{align}
where $\lVert\cdot\rVert_\textrm{1}$ is the $\ell_1$ norm and $N$ is the number of STFT frames.
The MSS loss is defined as the average error across each of the $M$ resolutions, i.e.,
\begin{align}
    \mathcal{L}_{\text{MSS}}(h, \hat{h})  = \frac{1}{M}\sum_{m=1}^M (\mathcal{L}_{\text{SC}}(h, \hat{h}) + \mathcal{L}_{\text{SM}}(h, \hat{h}))\,.
\end{align}

To achieve optimal performance, one must select the right frame size, window type, and hop size, as it is known, based on systematic analysis, that different hyperparameter configurations affect loss \cite{schwar2023multi}. However, Steinmetz and Reiss \cite{steinmetz2020auraloss} showed that randomly selecting these parameters each time the loss is computed can improve robustness. In our study, we use their default values  \cite{steinmetz2020auraloss}.

In addition to the MSS loss, we compute the error-to-signal ratio, ESR, defined as the squared error normalized by the energy of the target RIR \cite{wright2019real}, i.e.,
\begin{align}
    \mathcal{L}_{\textrm{ESR}}(h, \hat{h})  = \frac{\sum_{t_{\textrm{mix}}}^L\lvert h(t) - \hat{h}(t)\rvert^2}{\sum_{t_{\textrm{mix}}}^L\lvert h(t)\rvert^2} .
\end{align}
Unlike the other metrics, $\mathcal{L}_{\textrm{ESR}}$ incorporates the phase information of the analyzed RIRs. We considered this aspect as worthy of investigation, driven by the expectation that when RIRs are measured at identical microphone positions, phase differences will likely be lower in comparison to RIR pairs measured at distinct locations within the room.

\subsection{Reverberation condition differences}\label{subsec:study1}

First, we assess how the metrics respond to variations in the room's absorptive characteristics. We segmented the dataset into 11 partitions, determined by the number of reflective panels. For each partition, we randomly selected a subset of 25 RIRs, with 5 RIRs per microphone position. We apply the metrics to all possible pairs within and across the subsets. The median values, shown in Fig.~\ref{fig:med_loss_nclosed}, are computed after normalizing the data of each metric, ensuring zero mean and a unitary standard deviation. The color bar limits are set to the minimum and maximum median values among the four plots. Labels of the Y-axis refer to the reference RIR $h(t)$, while the X-axis refers to $\hat{h}(t)$. 

Among the metrics, $\mathcal{L}_{\textrm{ESR}}$ exhibits the most uniform distribution, whereas the other metrics display a decrease in the distance towards the diagonal, when RIRs from a partition are compared against themselves. Compared to $\mathcal{L}_{\textrm{MSS}}$, $\mathcal{L}_{\textrm{PC}}$ shows larger variation of the median towards the diagonal. Both $\mathcal{L}_{\textrm{MSS}}$ and $\mathcal{L}_{\textrm{PC}}$ show reduced sensitivity to changes in highly reverberant conditions (bottom right). Conversely, the lowest median values of $\mathcal{L}_{\textrm{EDC}}$ are distributed towards less reverberant conditions (top left). Furthermore, $\mathcal{L}_{\textrm{EDC}}$ yields more significant differences when comparing RIRs against a highly reverberant RIR. 

To be integrated as loss functions in an ML framework, the metrics must show a smooth decrease toward a minimum value, reflecting the acoustical configurations of the target RIR. To verify whether this is the case for the analyzed metrics, we selected a target RIR, which was measured with 20 reflective panels. We then calculated the distance between the target RIR and 50 randomly chosen RIRs for every number of reflective panels between 1--54, while ensuring the microphone position remained consistent with that of the target RIR. 

\begin{figure}[t!]
\centerline{\includegraphics[trim={1cm 0cm 2cm 0cm},width=0.49\textwidth]{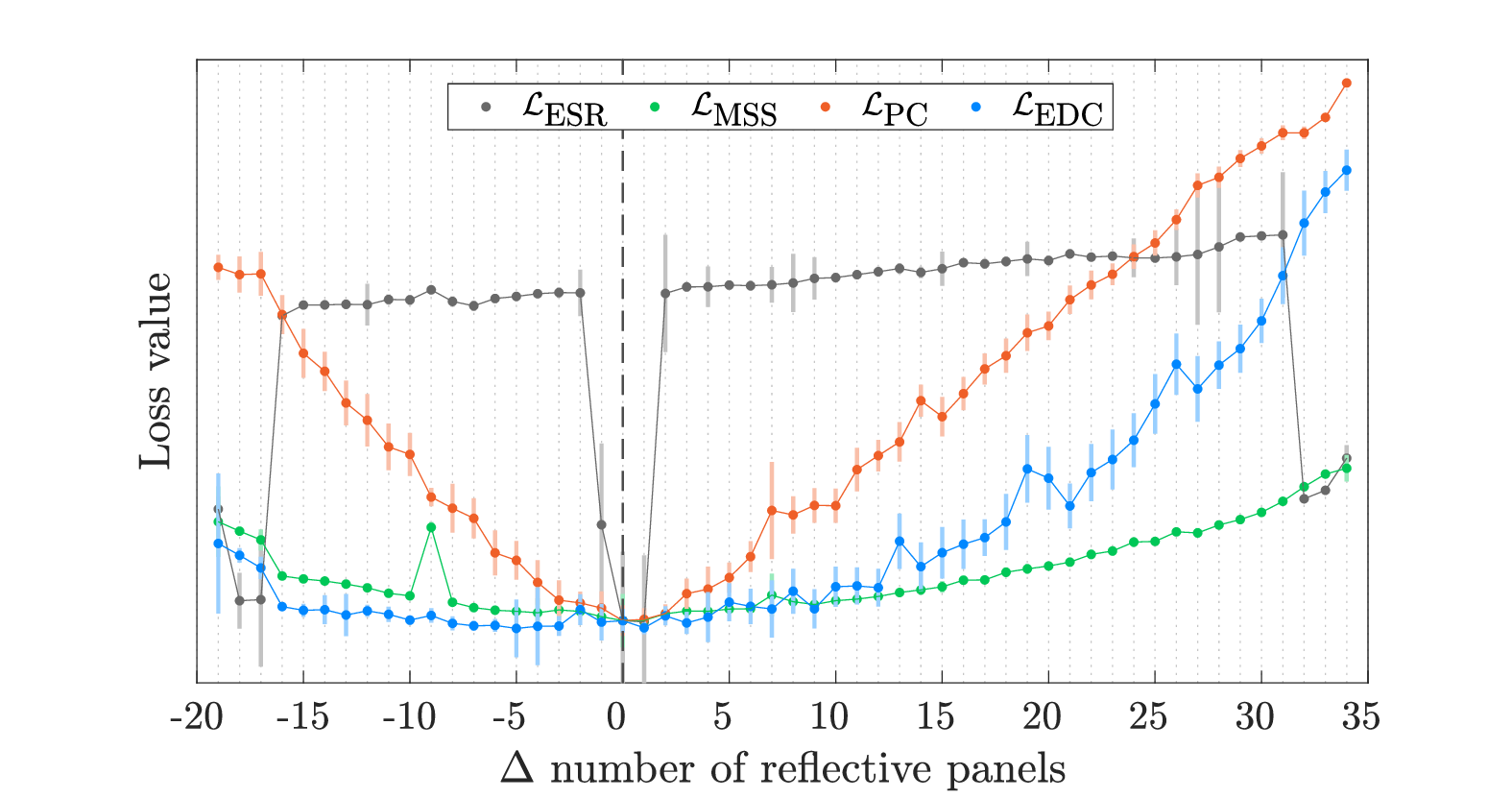}}
% \vspace{-1mm}
\caption{\textit{Evolution of metrics on gradual differences $\Delta$ in the number of panels set to a reflective position. Medians and stardard deviations are marked with dots and vertical lines, respectively. The dashed line indicates the reference RIR's configuration, measured with 20 reflective panels. $\mathcal{L}_{\textrm{PC}}$ shows the smoothest and most symmetric behavior.}}
\label{fig:smooth}
\end{figure}

Figure~\ref{fig:smooth} shows the median and standard deviation of the distance values, marked respectively with points and vertical dashes. The metrics have been normalized to their minimum and maximum values. Among them, $\mathcal{L}_{\textrm{PC}}$ and $\mathcal{L}_{\textrm{MSS}}$ exhibit the highest symmetry around the target reverberation configuration. $\mathcal{L}_{\textrm{PC}}$  demonstrates a pronounced but gradual rise towards larger differences $\Delta$. Similarly, $\mathcal{L}_{\textrm{EDC}}$ also shows a gradual increase, but it flattens when it compares the target to more absorbent configurations (on the left). $\mathcal{L}_{\textrm{ESR}}$ only detects similarities within very similar configurations and with a large standard deviation. Additional analysis would be required to understand why $\mathcal{L}_{\textrm{ESR}}$ decrases for large $\Delta$s. Overall, $\mathcal{L}_{\textrm{PC}}$ shows the most desirable behavior.

\begin{figure*}[t!]
\centerline{\includegraphics[trim={5cm 2cm 3cm 4cm}, width=\textwidth]{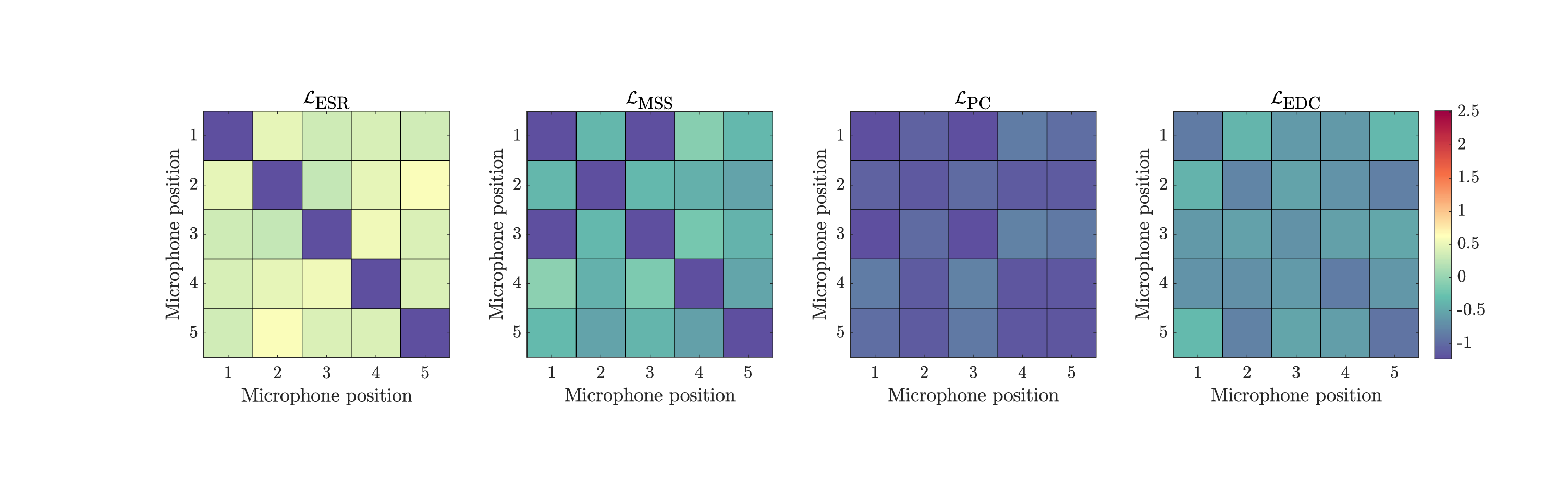}}
% \vspace{-1mm}
\caption{\textit{Median values of the similarity metric distribution for each pair of microphone positions. The values correspond to Fig.~\ref{fig:med_loss_nclosed} for the number of reflective panels in the range 35 to 49 for both $h(t)$ and $\hat{h}(t)$. Labels of the Y-axis refer to $h(t)$, while the X-axis refers to $\hat{h}(t)$. $\mathcal{L}_{\textrm{PC}}$ returns the smallest median values, which is a desired characteristic.}}
\label{fig:med_loss_mic}
\end{figure*}

\subsection{Receiver location differences}

From the metric values computed in Sec.~\ref{subsec:study1}, we isolate those relative to subsets with a number of reflective panels from 35--49, for both $h(t)$ and $\hat{h}(t)$. We then plot the median of the metrics for each pair of microphone positions in Fig.~\ref{fig:med_loss_mic}. Among the metrics, $\mathcal{L}_{\textrm{ESR}}$ and $\mathcal{L}_{\textrm{MSS}}$ exhibit the least homogeneous distribution. Both metrics display sharp minima for RIRs measured with the same microphone position. Additionally, $\mathcal{L}_{\textrm{MSS}}$ fails to detect differences between RIRs measured at microphone position 3 and those measured at position 1. Conversely, $\mathcal{L}_{\textrm{PC}}$ and $\mathcal{L}_{\textrm{EDC}}$ demonstrate a more uniformly distributed set of medians, with the former exhibiting lower values.

% ================= DISCUSSION ================= %
\section{Discussion}\label{sec:discussion}
This section delves into the results of the objective evaluation tests outlined earlier.
Similarly to Sec.~\ref{sec:eval}, we organize the discussion according to individual test types. 

\subsection{Reverberation condition differences}
Considering that all metrics in Fig.~\ref{fig:med_loss_nclosed} were normalized to equal standard deviation, $\mathcal{L}_{\textrm{PC}}$ and $\mathcal{L}_{\textrm{EDC}}$ display greater variation, indicating that they better capture differences between partitions and are more robust to outliers. $\mathcal{L}_{\textrm{EDC}}$ exhibits a bias towards more reverberant conditions, evidenced by higher values when either the $h(t)$ or $\hat{h}(t)$ belong to the most reverberant configurations. This bias might be an indicator of its efficacy in capturing the behavior of the RT in Arni \cite[Fig.~4]{prawda2022calibrating} which, especially at low-frequency bands,  increases almost exponentially with the number of panels in a reflective position, leading to larger differences when a RIR is compared against RIRs that belong to partition 50--55.

Regarding integration into ML frameworks, Fig.~\ref{fig:smooth} suggests that $\mathcal{L}_{\textrm{PC}}$ offers the most optimal profile among the analyzed metrics. The flatness of the $\mathcal{L}_{\textrm{ESR}}$ function in Fig.~\ref{fig:smooth} suggests that this loss is unsuitable as a reverb similarity metric in ML applications. $\mathcal{L}_{\textrm{MSS}}$ exhibits less noise, even though its trough is not as pronounced as those of $\mathcal{L}_{\textrm{PC}}$ and $\mathcal{L}_{\textrm{EDC}}$, which can potentially lead to slower learning rates.

\subsection{Receiver location differences}
Fig.~\ref{fig:med_loss_mic} shows that  $\mathcal{L}_{\textrm{PC}}$ and $\mathcal{L}_{\textrm{EDC}}$ exhibit more generalization across microphone positions than $\mathcal{L}_{\textrm{ESR}}$ and $\mathcal{L}_{\textrm{MSS}}$, a desirable behavior when evaluating late reverberation similarity. Conversely, $\mathcal{L}_{\textrm{MSS}}$ and $\mathcal{L}_{\textrm{ESR}}$ appear overly sensitive to minor variations, failing to capture the convergence of statistical properties in the late reverb. Furthermore, $\mathcal{L}_{\textrm{MSS}}$, and to a much lesser extent also $\mathcal{L}_{\textrm{PC}}$, appear to exhibit a greater degree of confusion between positions 1 and 3, despite their distinct locations \cite[Fig.~3]{prawda2022calibrating}. The $\mathcal{L}_{\textrm{ESR}}$ showed the poorest performance, indicating that similarity was only detected between RIRs measured at the same microphone position. This highlights the significance of time-frequency energy representation, and in particular of averaged quantities. These results suggest that the windowing operation carried out by STFT alone may not be sufficient to make $\mathcal{L}_{\textrm{MSS}}$ robust to minor and negligible noise-like differences.

\section{Conclusion}\label{sec:conclusion}
Two novel metrics for late reverberation similarity are proposed, one based on averaged power convergence ($\mathcal{L}_{\textrm{PC}}$) and the other on frequency-band energy decay ($\mathcal{L}_{\textrm{EDC}}$). To validate their performance, we used a dataset of RIRs collected in a variable acoustics room, enabling us to analyze fine changes between reverberation conditions. The metrics were compared to a time-domain error ratio and a popular multi-scale spectral loss. 

The proposed metrics show more robustness to changes in microphone position than baseline methods, which suggests they are more sensitive toward the acoustic features of reverberation rather than sample-wise differences. Objective tests showed that $\mathcal{L}_{\textrm{PC}}$ was better at capturing gradual changes in reverberation conditions, while the values of the baseline metrics exhibited the most uniform distribution across test cases. Moreover, $\mathcal{L}_{\textrm{PC}}$ displayed the most optimal decay towards the global minimum, indicating its potential as a loss function in ML applications. 

This study suggests how metrics can be optimized for reverberation, with the support of a well-documented and comprehensive dataset of RIRs. Future work includes a listening test to assess the correlation between subjective scores and the objective results presented in this study. 
% include analyzing various configurations of hyperparameters, and exploring different non-linear time-frequency scales. 
 % With the aim of applying these metrics as loss functions within a ML framework, thier performance have to be validated in a training pipeline. 

\bibliographystyle{unsrt}
\bibliography{bibliography}

\end{document}